\documentclass[aps,showpacs,nofootinbib,superscriptaddress,twocolumn]{revtex4}

\usepackage{graphicx}
\usepackage{bm}
\usepackage{amsmath}
\usepackage{amssymb}
\usepackage{hyperref}

\newcommand{\xbj}{x}
\newcommand{\qslash}{\kern 0.2 em n\kern -0.50em /}
\newcommand{\nslash}{\kern 0.2 em n\kern -0.50em /}
\newcommand{\kslash}{\kern 0.2 em k\kern -0.45em /}
\newcommand{\lslash}{\kern 0.2 em l\kern -0.50em /}
\newcommand{\pslash}{\kern 0.2 em p\kern -0.50em /}
\newcommand{\Sslash}{\kern 0.2 em S\kern -0.50em /}
\newcommand{\Pslash}{\kern 0.2 em P\kern -0.50em /}
\newcommand{\Dslash}{\kern 0.2 em D\kern -0.65em /\kern 0.15em}

\newcommand{\bp}{\boldsymbol{p}_T}
\newcommand{\bP}{\boldsymbol{P}_T}
\newcommand{\bk}{\boldsymbol{k}_T}

\newcommand{\eps}{\epsilon}
\newcommand{\ssh}{\!\!\!/}
\newcommand{\Tr}{\operatorname*{Tr}\nolimits}

\newcommand{\ph}{\phi_h}

\begin{document}

\title{ On the $\cos\phi_h$ asymmetry of electroproduction of pions in double longitudinally polarized process}

\author{Wenjuan Mao}\affiliation{Department of Physics, Southeast University, Nanjing
211189, China}
\affiliation{School of Physics and State Key Laboratory of Nuclear Physics and Technology, Peking University, Beijing
100871, China}
\author{Xiaoyu Wang}\affiliation{Department of Physics, Southeast University, Nanjing
211189, China}
\author{Xiaozhen Du}\affiliation{School of Physics and State Key Laboratory of Nuclear Physics and Technology, Peking University, Beijing
100871, China}
\author{Zhun Lu}\email{zhunlu@seu.edu.cn}\affiliation{Department of Physics, Southeast University, Nanjing
211189, China}
\author{Bo-Qiang Ma}\email{mabq@pku.edu.cn}\affiliation{School of Physics and State Key Laboratory of Nuclear Physics and Technology, Peking University, Beijing
100871, China}
\affiliation{Collaborative Innovation Center of Quantum Matter, Beijing, China}
\affiliation{Center for High Energy Physics, Peking University, Beijing 100871, China}

\begin{abstract}
We study the $\cos{\phi_h}$ azimuthal asymmetry in double polarized semi-inclusive pion production by considering dynamical twist-3 effects.
In particular, we evaluate the role of the transverse momentum dependent distributions $e_L(x, \bm k_T^2)$ and $g_L^\perp(x, \bm k_T^2)$ on the asymmetry.
Using two different sets of spectator model results for these distributions, we predict the $\cos{\phi_h}$ asymmetry of $\pi^+$, $\pi^-$, and $\pi^0$ at the kinematic configuration available at CLAS, HERMES and COMPASS.
Our estimates show that the asymmetries are positive for all the pions and could be accessed by CLAS and HERMES.
We also find that $g_L^\perp$ gives the dominant contribution to the $\cos\phi_h$ asymmetry, while the contribution of $e_L$ is almost negligible.

\end{abstract}

\pacs{12.39.-x, 13.60.-r, 13.88.+e}

\maketitle

\section{Introduction}

Nowadays it is very clear that a better understanding on the nucleon structure can be achieved if one goes beyond the collinear picture to take into account the transverse degree of freedom of partons.
Early investigation~\cite{Cahn} on the unpolarized semi-inclusive deep inelastic scattering (SIDIS) demonstrated that the intrinsic transverse motion of quarks can give rise to a $\cos\phi_h$ asymmetric distribution of the final state hadron.
This mechanism, usually called as the Cahn effect, provides a useful tool to probe the partonic transverse momentum which is still less understood so far, although the effect appears kinematically at the subleading order of inverse hard scale.
The same idea was then explored further by several experimental and theoretical  studies~\cite{Ashman:1991cj,Aubert:1983cz,Arneodo:zpc34,Adams:1993hs,Mkrtchyan:2007sr,Osipenko:2008aa,Airapetian:2012yg,Adolph:2014npb,
Anselmino:2005prd,Schweitzer:2010tt,Boglione:2011} to give constraints on the transverse structure of nucleon, i.e., the effect was applied~\cite{Anselmino:2005prd,Schweitzer:2010tt} to extract the average values of the intrinsic transverse momenta of quarks inside the nucleon from SIDIS~\cite{Ashman:1991cj,Aubert:1983cz,Arneodo:zpc34,Adams:1993hs} or Drell-Yan data~\cite{Conway:1989fs}.

Recently, in a new study~\cite{Anselmino:2006yc} the Cahn effect was extended to the case of double longitudinally polarized SIDIS, prediciting a similar $\cos\phi_h$ azmithual asymmetry
that originates from the transverse momentum dependent (TMD) helicity distribution function $g_{1L}(x,\bm k_T^2)$.
Due to the positive value of $g_{1L}^{u}(x,\bm k_T^2)$ and $u$ quark dominance, the asymmetry based on the Cahn effect was found to be negative for both the charged and neutral pions in the case of proton target.

In this work, we will study the $\cos\phi_h$ azimuthal asymmetry in double longitudinally polarized SIDIS in an alternative approach, that is, to employ dynamical twist-3 effects.
As shown in Ref.~\cite{Bacchetta:0611265}, the polarized structure function $F_{LL}^{\cos\phi_h}$ that associated with the $\cos\phi_h$ asymmetry can be expressed in terms of the twist-3 TMD distribution/fragmentation function combined with the twist-2 fragmentation/distribution function.
Particularly, two twist-3 TMD distributions appear in the convolutions: the T-even distribution $g_L^\perp(x,\bm k_T^2)$ and the T-odd distribution $e_L(x,\bm k_T^2)$.
The former one can be decomposed into the following form via the equation of motion relation~\cite{Mulders:1995dh}
\begin{align}
x\,g_L^\perp = x\,\tilde{g}_L^\perp +g_{1L} + {m\over M} h^\perp_{1L}.\label{eq:emr}
\end{align}
Taking the component $g_{1L}$ from $x\,g_L^\perp$ in the above equation is equivalent to adopting the Cahn effect.
In this study we will consider the effect of the entire twist-3 distribution $g_L^\perp$.
In addition, we also take into account the contribution of $e_L$ coupled with the Collins fragmentation function~\cite{Collins:1993npb}.
We calculate these two TMD distributions of valence quarks inside the proton by employing the spectator diquark model
and predict the corresponding $\cos\phi_h$ asymmetry for charged and neutral pions at the kinematics of JLab, HERMES and COMPASS.
We note that sizable dynamical twist-3 effects may also appear in other processes which involve different polarizations of the lepton beam and the nucleon target~\cite{Airapetian:2005jc, hermes07,Alekseev:2010dm,Aghasyan:2011ha,Gohn:2014zbz,Parsamyan:2014uda,Metz:2004je,Bacchetta:2004zf,
Mao:2012dk,Song:2013sja,Song:2014sja,Lu:2014fva,Mao:2014aoa}.

\section{Model Calculations on the twist-3 TMD distributions $g_L^\perp$  and $e_L$ }
\label{functions}

In this section, we briefly present our calculation on the distributions $g_L^\perp(x,\bm k_T^2)$  and $e_L(x,\bm k_T^2)$ using the spectator model~\cite{Jakob:1997npa,Brodsky:2000ii,Brodsky:2002cx,jy02,Bacchetta:plb578,Gamberg:2006ru,Bacchetta:2008af,Lu:2012gu}
We will consider the contributions from both the scalar diquark and the vector diquark.
In the case of vector diquark components, we use two approaches for comparison.
The main differences between them are the form for the vector diquark propagator, as well as the flavor separation for $u$ and $d$ valence quarks.

The gauge-invariant quark-quark correlator for a longitudinally polarized nucleon in SIDIS reads:
\begin{align}
\Phi(x,\bm k_T)&=\int {d\xi^- d^2\bm{\xi}_T\over (2\pi)^3}e^{ik\cdot\xi}
\langle PS_L|\bar{\psi}_j(0)\mathcal{L}^{\bm 0_T}[0^-,\infty^-]\nonumber\\
& \times \mathcal{L}^{\xi^-}[\bm 0_T,\bm \xi_T]\mathcal{L}^{\bm \xi_T}[\infty^-,\xi^-]\psi_i(\xi)|PS_L\rangle\,.
\label{eq:Phi}
\end{align}
Here the light-cone coordinate  $a=[a^-,a^+,\bm a_T]$ is employed, $\mathcal{L}$s are the gauge links ensuring the gauge invariance of the operator, and $k$ and $P$ are the momenta of the struck quark and the target nucleon, respectively.
The distributions  $g_L^\perp$ and $e_L$ thus can be obtained from the correlator~(\ref{eq:Phi}) using the traces~\cite{Bacchetta:0611265,Goeke:2005hb}
\begin{align}
S_L\frac{\eps_{T}^{\alpha\rho} k_{T \rho}^{}}{P^+} \,
  g_L^{\perp}(x,\bm{k}_{T}^{2})& =
-\frac{1}{2}\Tr[\Phi\gamma^{\alpha}\gamma_5]\,, \label{phitr2}\\
S_L\frac{M}{P^+} e_L(x,\bm k_T^2)   & =
\frac{1}{2}\Tr[\Phi i\gamma_5]\,.\label{eq:eltr}
\end{align}

First we consider the contribution from the scalar diquark.
In the lowest order, the correlator may be calculated by suppressing the gauge link in the operator.
After some algebra, we arrive at the expression for the correlator:
\begin{align}
\Phi^{(0)}_s(x,\bk)&\equiv \frac{N_s^2(1-x)^3}{32 \pi^3 P^+}\frac{\left[ (k\ssh +m)\gamma_5 S\ssh (P\ssh +M) (k\ssh +m)\right]}{(\bk^2+L_s^2)^4}\,. \label{lophis}
\end{align}
Here $N_s$ is the normalization constant, and the notation $L_s^2$ has the form
\begin{align}
L_s^2=(1-x)\Lambda_{s}^2 +x M_{s}^2-x(1-x)M^2,\label{Ls2}
\end{align}
with $\Lambda_s$ being the cutoff parameter for the quark momentum and $M_s$ the scalar diquark mass.

The lowest order result in Eq.~(\ref{lophis}) can be used to calculate T-even distributions.
However, it leads to a vanishing result for T-odd distributions.
In order to yield a nonzero contribution from the correlator, one has to consider the effect of the gauge links~\cite{Brodsky:2002cx,jy02,Collins:2002plb}, or the rescattering between the struck quark and the spectator diquark.
Here we expand the gange-links to the first nontrivial order, which corresponds to the one gluon exchange approximation. At this order, the correlator has the form:
\begin{align}
 \Phi_s^{(1)}
(x,\bm k_T)
&\equiv
i g N_{s}^2 {(1-x)^2\over 64\pi^3 (P^+)^2}\frac{-i\Gamma^{+}_s}{(\bm{k}_T^2+L_s^2)^2}\nonumber \\
\hspace{-1cm}&\times \int {d^2 \bm q_T\over (2\pi)^2}
{ \left[(\kslash -q\ssh+m)\gamma_5 S\ssh (\Pslash+M)(\kslash +m)\right]
\over \bm q_{T}^2  \left[(\bm{k}_T-\bm{q}_T)^2+L_s^2\right]^2}
\,, \label{phis1}
 \end{align}
with $q^+=0$, and $\Gamma_s^\mu$ being the vertex between the gluon and the scalar diquark:
\begin{align}
 \Gamma_s^\mu &= ig (2P-2k+q)^\mu \,.\label{Gammas}
\end{align}

Substituting (\ref{phis1}) into (\ref{eq:eltr}) and (\ref{lophis}) into (\ref{phitr2}), we obtain the following expressions for $g_L^\perp$ and $e_L$ from the scalar diquark component:
\begin{align}
g_L^{\perp s}(x,\bk^2)&=-\frac{N_s^2(1-x)^2}{16\pi^3}\frac{(1-x)^2 M^2-M_s^2-\bk^2}{(\bk^2+L_s^2)^4}\,,\label{glperps} \\
e_L^{s}(x,\bk^2)&=C_F\alpha_s\frac{{N_s}^2(1-x)^2}{32\pi^3} \frac{(x+\frac{m}{M})(L_s^2-\bk^2)}{L_s^2(L_s^2+\bk^2)^3}\,.
 \label{els}
\end{align}
We note that the result in (\ref{glperps}) has already been given in Ref.~\cite{Jakob:1997npa}.

The correlator contributed by the vector diquark can be obtained in the similar way which was applied to calculate $\Phi_s^{(0)}$ and $\phi_s^{(1)}$.
Here we cast the expressions for the correlator from the vector diquark component at the lowest order:
\begin{align}
\Phi^{(0)}_{v}(x,\bk)&\equiv \frac{N_v^2(1-x)^3}{64 \pi^3 P^+}d_{\mu\nu}(P-k)\nonumber\\
&\times \frac{\left[(k\ssh +m)\gamma^{\mu}\gamma_5 S\ssh(M-P\ssh )\gamma^{\nu}(k\ssh+m)\right]}{(\bk^2+L_v^2)^4}\,, \label{lophiv}
\end{align}
and at the one-loop level:
\begin{align}
 \Phi^{(1)}_{v}
(x,\bm k_T)
&\equiv
i g N_v^2 {(1-x)^2\over 128\pi^3 (P^+)^2}{1\over (\bm{k}_T^2+L_v^2)^2}\nonumber\\
&\times\int {d^2 \bm q_T\over (2\pi)^2} \,
 d_{\rho\alpha}(P-k)\, (-i\Gamma^{+,\alpha\beta}) \nonumber\\
 &\times d_{\sigma\beta}(P-k+q) \nonumber\\
&\times{ \left[(\kslash -q\ssh+m) \gamma^\sigma \gamma_5 S\ssh(M-\Pslash)\gamma^\rho (\kslash +m)\right]
\over \bm q_T^2  \left[(\bm{k}_T-\bm{q}_T)^2+L_v^2\right]^2}\,,
\label{phia1}
\end{align}
respectively.
In Eqs.~(\ref{lophiv}) and (\ref{phia1}), we have used $d_{\mu\nu}$ to denote the propagator of the vector diquark,  which corresponds to the sum of its polarization vectors.
Also, $\Gamma_v^{\mu,\alpha\beta}$ denotes the vertex between the gluon and the vector diquark
\begin{align}
 \Gamma_v^{\mu,\alpha\beta} &=  -i g [(2P-2k+q)^\mu g^{\alpha\beta}-(P-k+q)^{\alpha}g^{\mu\beta}\nonumber\\
 &-(P-k)^\beta g^{\mu\alpha}]\,. \label{Gamma}
\end{align}

In literature, different choices have been made for $d_{\mu\nu}$.
As shown in Ref.~\cite{Bacchetta:2008af}, different form of $d_{\mu\nu}$ generally leads to different result of the correlator.
In this work, we will consider two choices for $d_{\mu\nu}$ for comparison.
The first one has the form:
\begin{align}
 d^{\mu\nu}(k)  =& \,-g^{\mu\nu}\,+\, {k^\mu n_-^\nu
 \,+ \,k^\nu n_-^\mu\over k\cdot n_-}
  - \,{M_v^2 \over\left[k\cdot n_-\right]^2 }\,n_-^\mu n_-^\nu ,\label{d1}
\end{align}
which is motivated by the light-cone formalism~\cite{Brodsky:2000ii} for the vector diquarks.
Applying the propagator (\ref{d1}), we obtain the corresponding contributions to $g_L^\perp$ and $e_L$ from the axial-vector diquark component:
\begin{align}
g_L^{\perp v}(x,\bk^2){\big{|}}_{\textrm{Set I}}&=\frac{N_v^2(1-x)}{16\pi^3}\frac{(1-x)}{(\bk^2+L_v^2)^4}\nonumber\\
&\times\left[(m+xM)^2+(1-x)M^2-M_v^2+x\bk^2\right],\label{glperpv1} \\
e_L^{v}(x,\bk^2){\big{|}}_{\textrm{Set I}}&=0\,,\label{elv1}
\end{align}
and we denote them as the Set I results of $f^{v}$.

The second form for the vector diquark propagator employed in our calculation is
\begin{align}
d^{\mu\nu}(k) =& \,-g^{\mu\nu},
\label{d2}
\end{align}
which has been applied in Ref.~\cite{Bacchetta:plb578}.
Similarly, using (\ref{d2}) we obtain alternative expressions for $g_L^{\perp v}$ and $e_L^{v}$:
\begin{align}
e_L^{v}(x,\bk^2){\bigg{|}}_{\textrm{Set II}}&=C_F\alpha_s\frac{{N_v}^2(1-x)^2}{32\pi^3} \frac{(x+\frac{m}{M})(L_v^2-\bk^2)}{L_v^2(L_v^2+\bk^2)^3},\label{elv2}\\
g_L^{\perp v}(x,\bk^2){\bigg{|}}_{\textrm{Set II}}&=\frac{N_v^2(1-x)^2}{16\pi^3}\frac{(1-x)^2 M^2-M_v^2-\bk^2}{(\bk^2+L_v^2)^4},
\label{glperpv2}
\end{align}
which we denote as Set II results.

\begin{figure}
  \includegraphics[width=0.99\columnwidth]{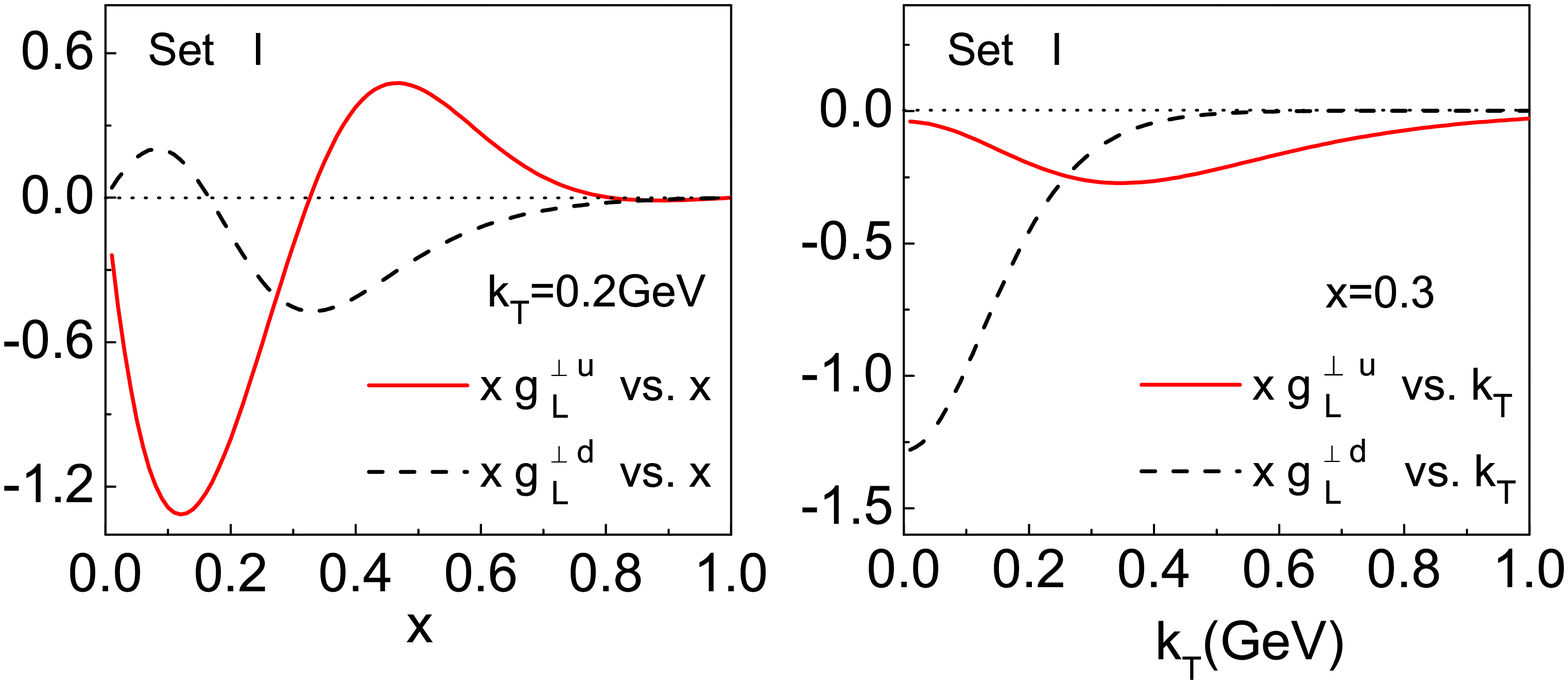}
  \caption{Model results for $x g_L^\perp(x,\bm k_T^2)$ in Set I.
  Left panel: the $x$ dependence of $x g_L^{\perp u}$ (solid line) and $x g_L^{\perp d}$ (dashed line) at $k_T=0.2\,\text{GeV}$;
  right panel: the $k_T$ dependence of $x g_L^{\perp u}$ (solid line) and $x g_L^{\perp d}$ (dashed line) at $x=0.3$.}\label{fig:glperp1}
\end{figure}

\begin{figure}
  \includegraphics[width=0.99\columnwidth]{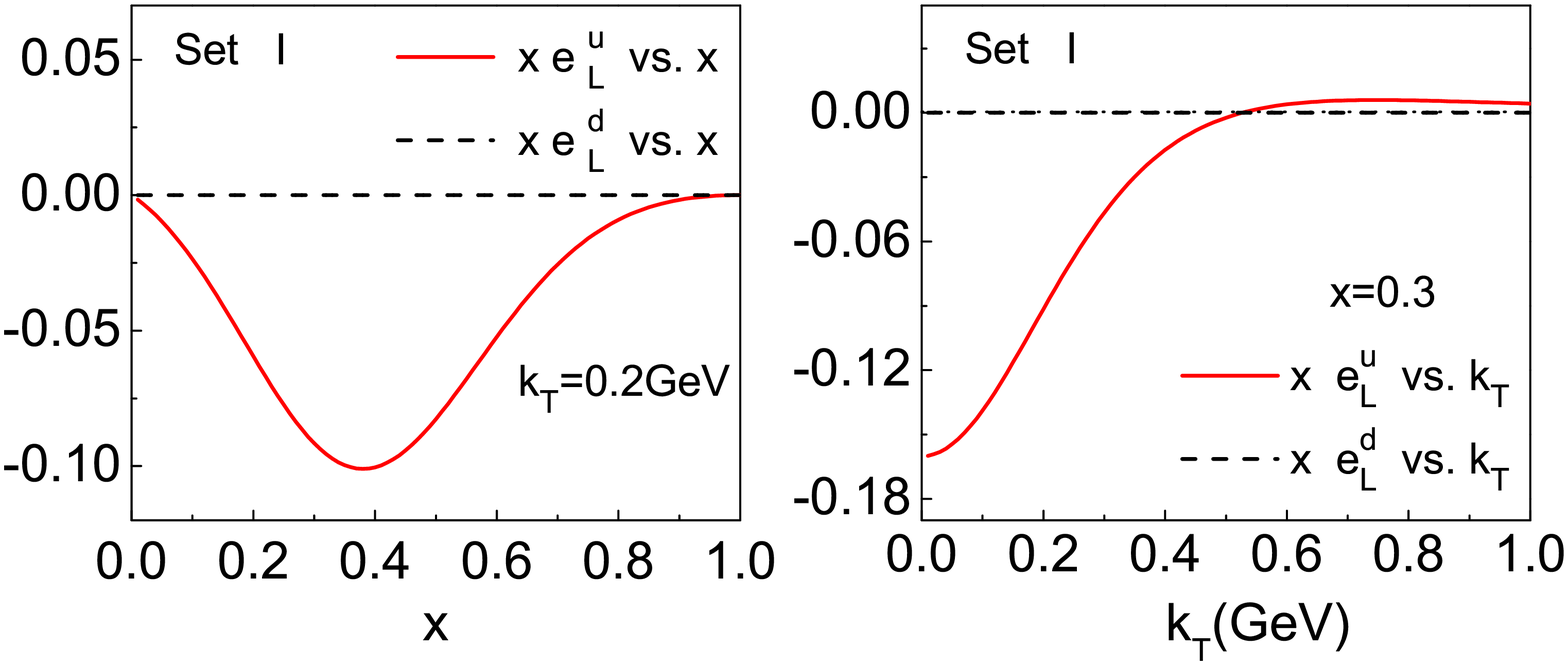}
  \caption{Similar to Fig.~\ref{fig:el1}, but for the model results of $x e_L^{u}$ (solid line) and $x e_L^{d}$ (dashed line).}\label{fig:el1}
\end{figure}

In order to obtain the flavor dependence of the TMD distributions, one should assign the relation between the quark flavors and the diquark types.
In Ref.~\cite{Bacchetta:2008af}, a general relation is introduced:
\begin{align}
f^u\big{|}_{\textrm{Set I}}=c_s^2 f^s + c_a^2 f^a,~~~~f^d\big{|}_{\textrm{Set I}}=c_{a^\prime}^2 f^{a^\prime}\,,\label{ud}
\end{align}
where $a$ and $a^\prime$ represent the vector isoscalar diquark $a(ud)$ and the vector isovector diquark $a(uu)$, respectively, and
$c_s$, $c_a$ and $c_{a^\prime}$ are the parameters of the model.
These parameters as well as the mass parameters (such as the diquark masses $M_{s/v}$, cut-off parameters $\Lambda_{s/v}$) are fitted from the ZEUS unpolarized parton distribution functions~\cite{zeus} and GRSV01 polarized parton distribution functions~\cite{grsv01}.
We combine Eqs. (\ref{glperps}), (\ref{els}), (\ref{glperpv1}), and (\ref{elv1}) to obtain the Set I distributions $e_L^q$ and $g_L^{\perp\,q}$ with $q = u$ and $d$.
For the strong coupling appearing in the expressions for $e_L^X$,
we choose $\alpha_s\approx 0.3$.
In the left panels of Figs.~\ref{fig:glperp1} and \ref{fig:el1}, we plot the $x$-dependence (at $k_T=0.2~\textrm{GeV}$) of the functions $g_L^{\perp q}(x,\bm k_T^2)$ and $e_L^q(x,\bm k_T^2)$ timed with $x$ for $q=u$ and $d$ quarks in Set I. We also plot the $k_T$-dependence (at $x=0.3$) of the distributions in the right panels of Figs.~\ref{fig:glperp1} and \ref{fig:el1}.

\begin{figure}
  \includegraphics[width=0.99\columnwidth]{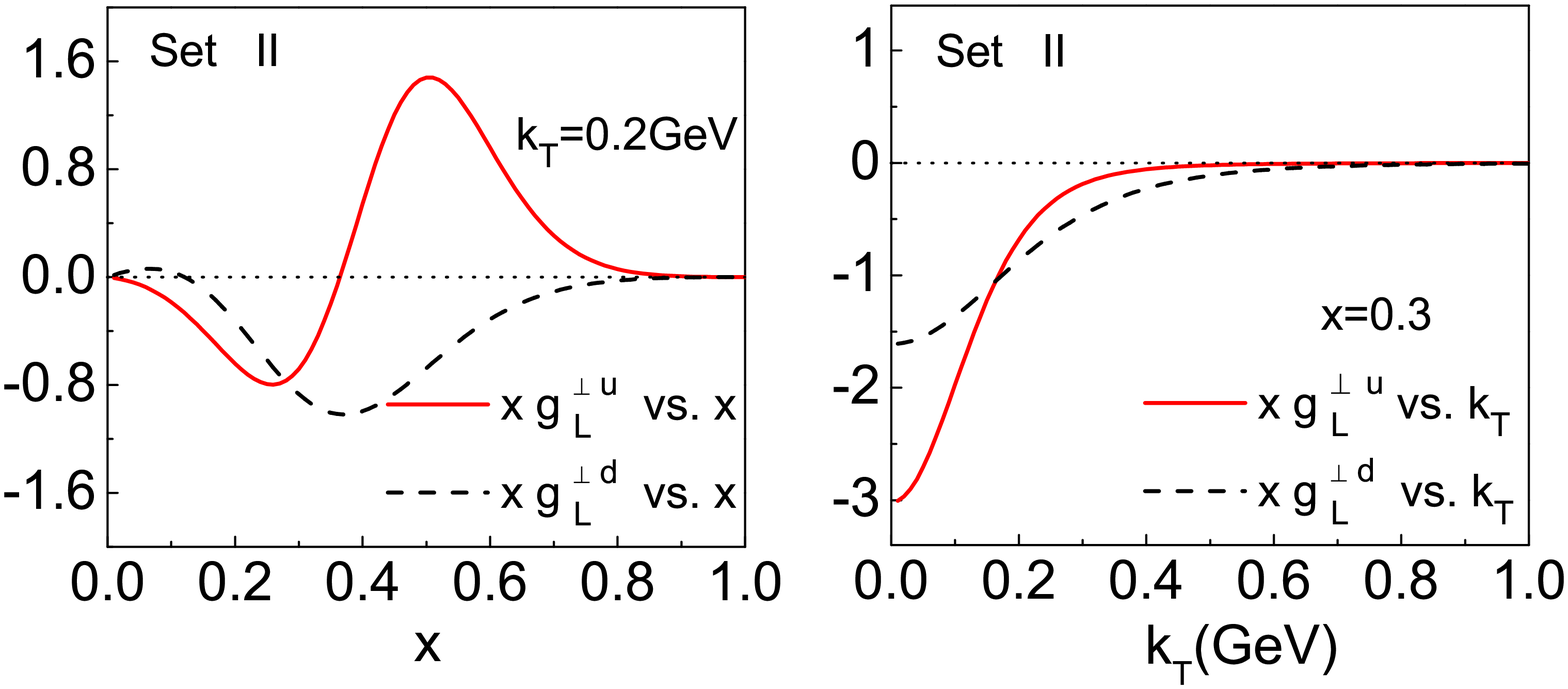}
  \caption{Model results for $x g_L^\perp(x,\bm k_T^2)$ in Set II. }\label{fig:glperp2}
\end{figure}

\begin{figure}
  \includegraphics[width=0.99\columnwidth]{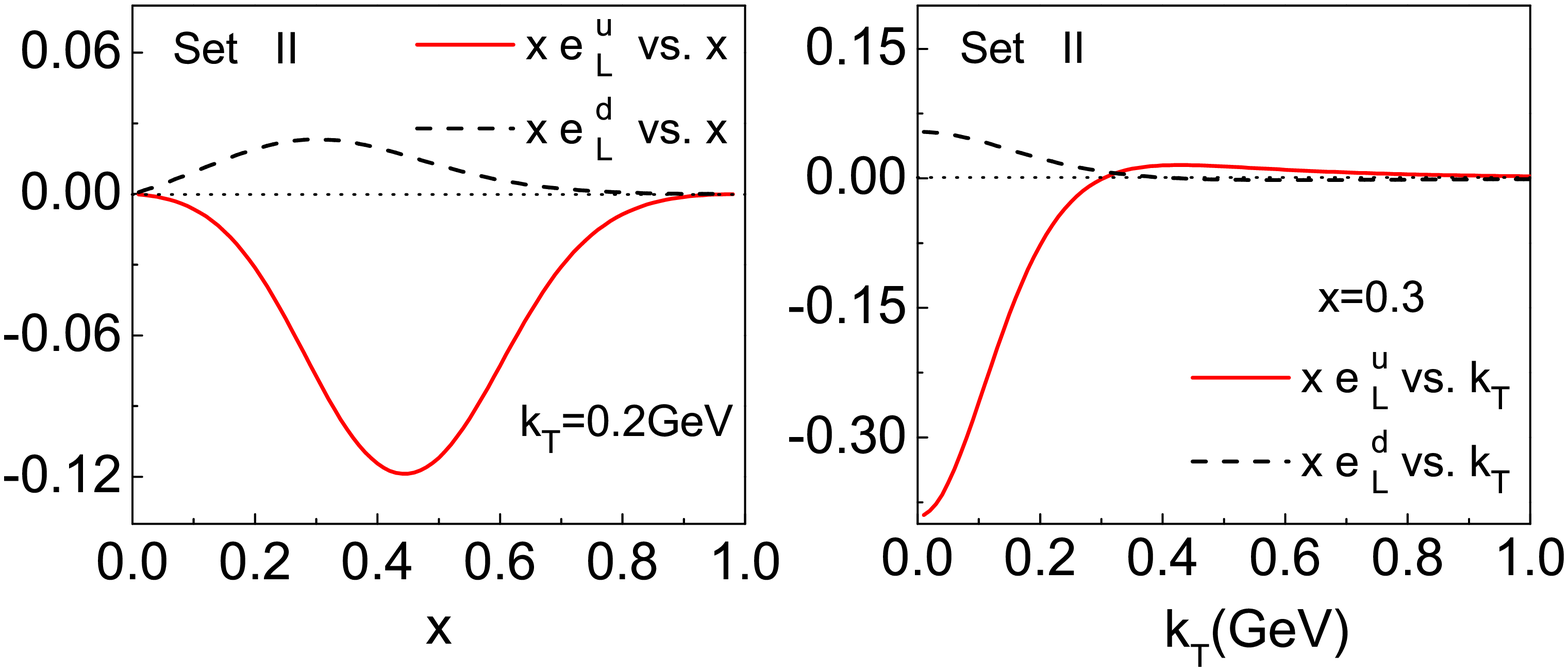}
  \caption{Model results for $x e_L(x,\bm k_T^2)$ in Set II.}\label{fig:el2}
\end{figure}

Different from Eq.~(\ref{ud}), another kind of flavor separation has been employed previously~\cite{Jakob:1997npa,Bacchetta:plb578}:
\begin{align}
f^u\big{|}_{\textrm{Set II}}=\frac{3}{2}f^s+\frac{1}{2} f^a,~~~~f^d\big{|}_{\textrm{Set II}}=f^{a^\prime}, \label{set2}
\end{align}
where the coefficients $3/2$, $1/2$ and $1$ in front of $f^X$ are obtained from the SU(4) spin-flavor symmetry of the proton wave function.
In this model, the mass parameters for different types of vector diquarks are the same, and we apply the values for the parameters from Ref.~\cite{Bacchetta:plb578}.
Using the relation (\ref{set2}), together with the expressions (\ref{glperps}), (\ref{els}), (\ref{glperpv2}) and (\ref{elv2}) , we obtain another set of TMD distributions, which we denote as Set II distributions.
The corresponding numerical results are plotted in Figs.~\ref{fig:glperp2} and \ref{fig:el2}.


Comparing Fig.~\ref{fig:glperp1} with Fig.~\ref{fig:glperp2} and Fig.~\ref{fig:el1} with Fig.~\ref{fig:el2}, we can see that the sizes of the TMD distributions in Set I are different from those in Set II.
In both sets, the signs of $e_L$ and $g_L^\perp$ turn to be negative in the specified kinematics ($x=0.3$ and $k_T=0.2$ GeV).
Also, the size of the T-even distribution $g_L^\perp$ is generally larger than that of the T-odd distribution $e_L$.
This is understandable since T-odd distributions are yielded from the higher order expansion of gauge link.
The distribution $g_L^\perp$ has also been calculated~\cite{Avakian:2010br} in the Bag model.
We find that the tendency of the $x$ dependence of $g_L^\perp$ in our calculation agrees with the result in Ref.~\cite{Avakian:2010br},
that is, there is a node in the intermediate $x$ region.
This behavior may be explained by the so called Lorentz invariance relation~\cite{Avakian:2010br,Mulders:1995dh,Teckentrup:2009tk}.
We also point out that our result for $e_L^q(x,\bm k_T^2)$ agrees with the time reversal constraint for distributions
$\int d^2 \bm k_T \, e_L^q(x,\bm k_T^2) =0$.

\section{Predictions on the $\cos\phi_h$ asymmetry for charged and neutral pions in polarized SIDIS}
\label{asy}

\begin{figure}
  \includegraphics[width=0.8\columnwidth]{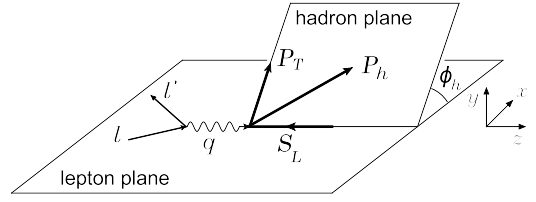}
 \caption {The kinematical configuration for longitudinally polarized SIDIS process.
 The initial and scattered leptonic momenta define the lepton plane ($x-z$ plane), while the momentum of the detected hadron together with the $z$ axis identifies the hadron production plane; the longitudinal spin of the nucleon is along the $-z$ axis.}
 \label{SIDISframe}
\end{figure}

In this section, we perform phenomenological analysis
on the $\cos\phi_h$ asymmetry for pions in SIDIS:
\begin{align}
l^\rightarrow(\ell) \, + \, p^\rightarrow (P) \, \rightarrow \, l' (\ell')
\, + \, h (P_h) \, + \, X (P_X)\,.
\label{sidis}
\end{align}
Here the arrow $\rightarrow$ denotes the longitudinally polarization of the beam or proton target, $\ell$ and $\ell'$ stand for the momenta of the incoming and outgoing leptons, and $P$ and $P_h$ denote the momenta of the target nucleon and the final-state hadron, respectively.
The kinematics of SIDIS can be expressed by the following invariant variables
\begin{align}
&x = \frac{Q^2}{2\,P\cdot q},~~~
y = \frac{P \cdot q}{P \cdot l},~~~
z = \frac{P \cdot P_h}{P\cdot q},~~~\gamma={2Mx\over Q},~~~\nonumber\\
&Q^2=-q^2, ~~~
s=(P+\ell)^2,~~~
W^2=(P+q)^2,~~~
\end{align}
with $q=\ell-\ell'$ the momentum of the virtual photon, and $W$ the invariant mass of the hadronic final state.
The reference frame adopted in this work is shown in Fig.~\ref{SIDISframe}, in which the momentum of the virtual photon is along the $z$ axis, and
the longitudinal polarization of the target is along the opposite direction of $z$ axis.
Thus, we will not consider the contribution from the transverse component of the polarization, which involves the TMD distribution $g_{1T}$.
In this frame, the transverse momentum of the final hadron with respect to the fragmenting quark is denoted by $\bP$, and the azimuthal angle of the hadron around the virtual photon is defined as $\phi_h$.

The differential cross section of SIDIS by scattering a longitudinally polarized lepton beam off a longitudinally polarized target can be expressed as~\cite{Bacchetta:0611265}
\begin{align}
\frac{d\sigma}{d\xbj dy\,dz dP^2_T d\ph} &=\frac{2\pi \alpha^2}{\xbj y Q^2}\frac{y^2}{2(1-\varepsilon)}
 \Bigl( 1+ \frac{\gamma^2}{2\xbj} \Bigr)
  \left\{ F_{UU} \right.\nonumber\\
  & + \left. S_{\parallel}\lambda_e \sqrt{2\varepsilon(1-\varepsilon)} \cos \phi_h \,\,F^{\cos \phi_h}_{LL}+\cdots \right\},\label{eq:cs}
\end{align}
here $F_{UU}$ and $F_{LL}^{\cos\phi_h}$ are the spin-averaged and spin-dependent structure functions, respectively, and the ratio of the longitudinal and transverse photon flux is defined by $\varepsilon=\frac{1-y-\gamma^2y^2/4}{1-y+y^2/2+\gamma^2y^2/4}$.
The ellipsis stands for the leading-twist double-spin asymmetry $A_{LL}$~\cite{Anselmino:2006yc,Lu:2012ez} which will not be analyzed in this work.

\begin{figure*}
  \includegraphics[width=0.67\columnwidth]{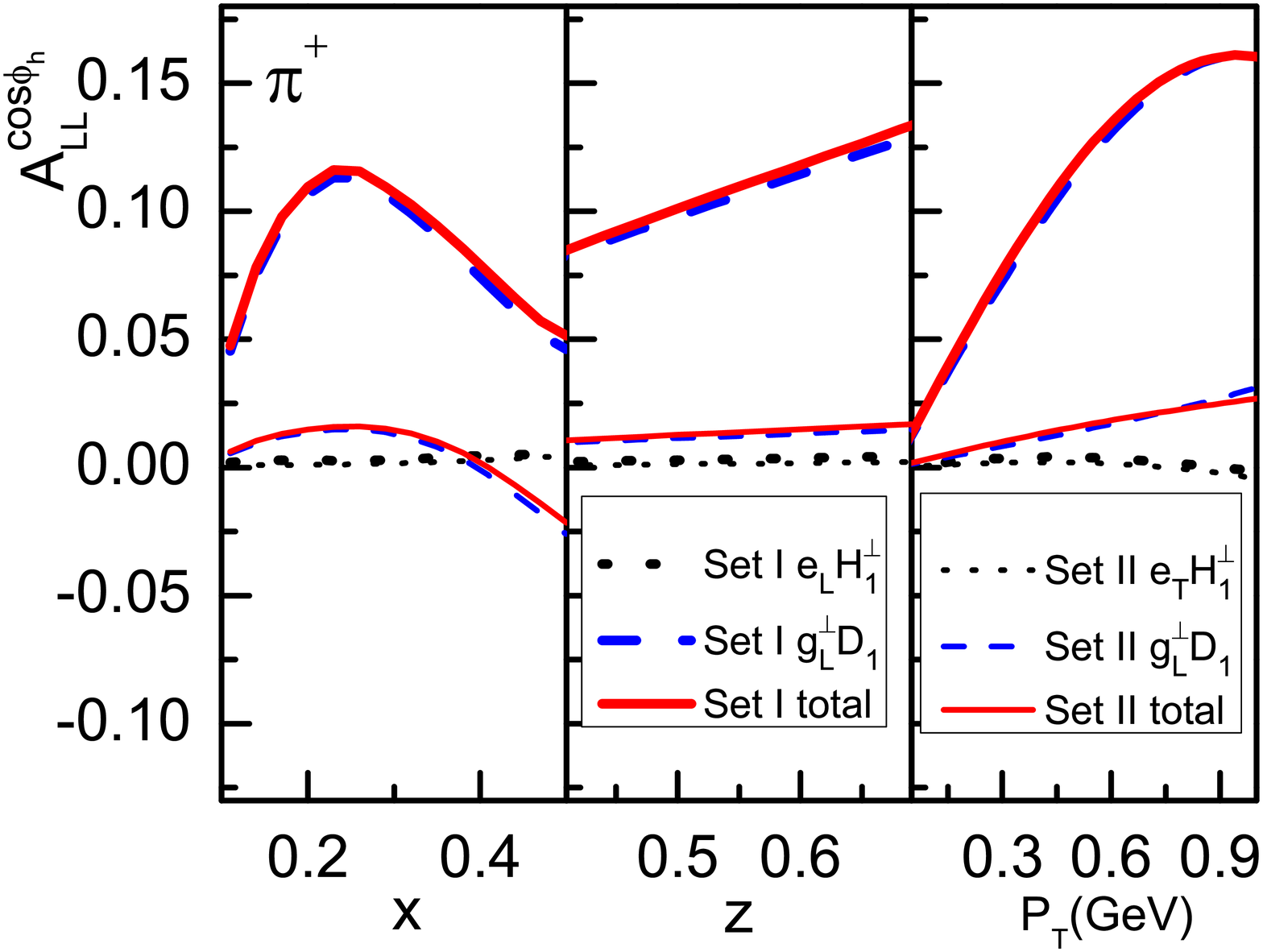}
  \includegraphics[width=0.67\columnwidth]{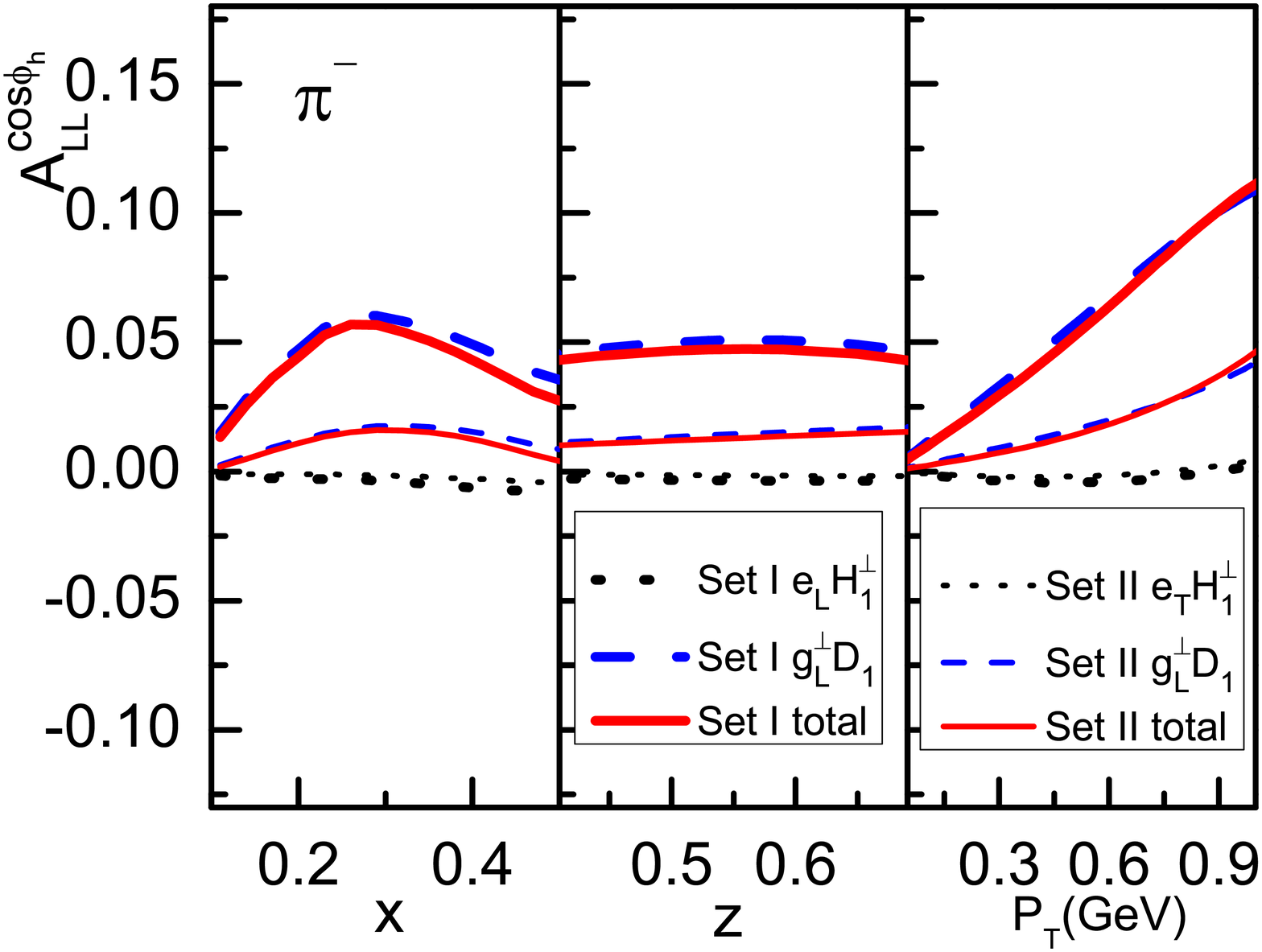}
  \includegraphics[width=0.67\columnwidth]{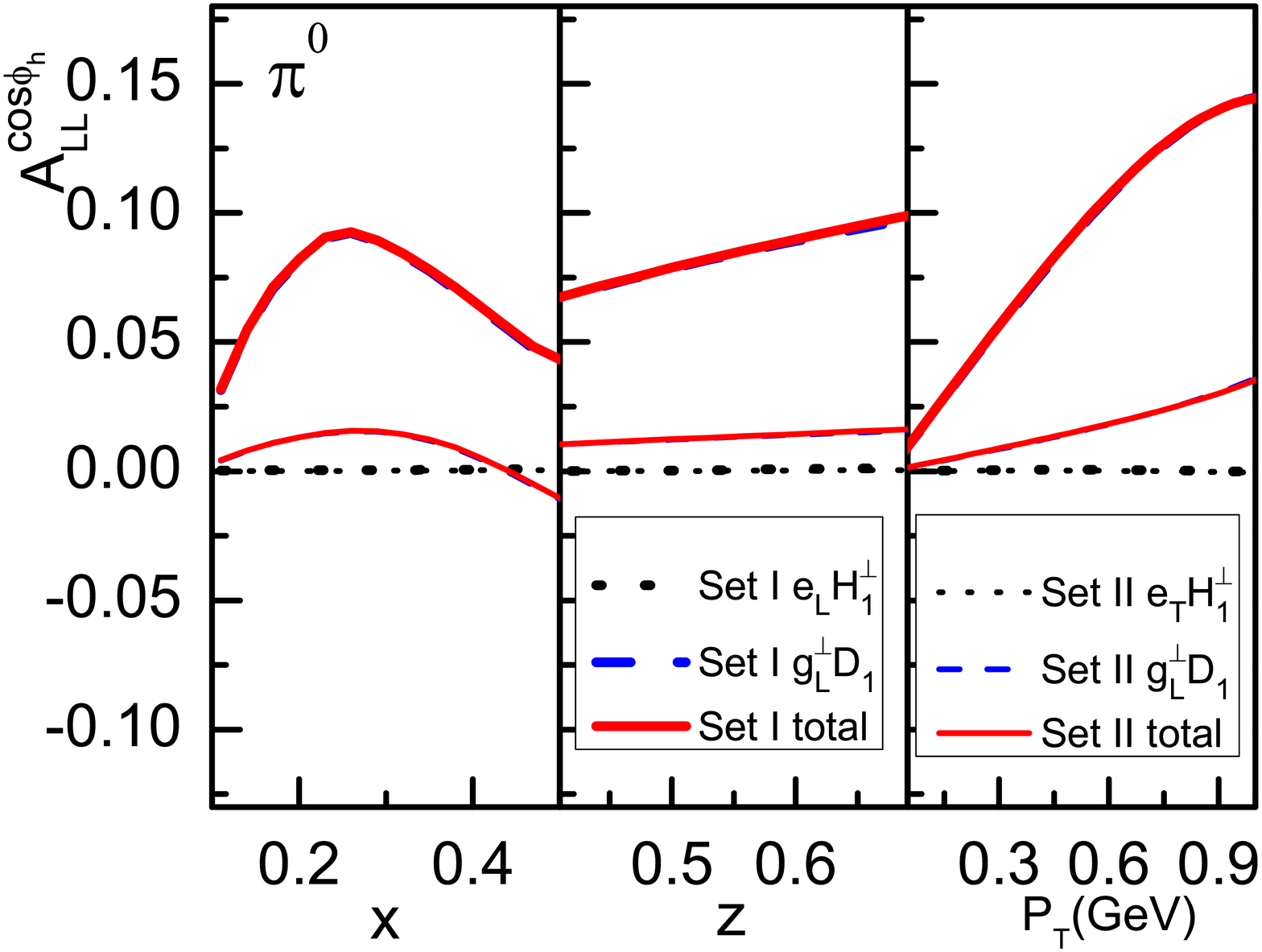}
  \caption{Prediction on $A_{\mathrm{LL}}^{\cos\phi_h}$ for $\pi^+$ (left panel), $\pi^-$ (central panel), and $\pi^0$ (right panel) vs $x$, $z$, and $P_T$ in SIDIS at CLAS off a proton target.
  The thick and thin curves correspond to the asymmetries calculated from the TMD distributions in Set I and Set II, respectively.
  The dotted and dashed curves represent the asymmetries from the $e_L H_1^\perp$ term and the $g_L^\perp D_1$ term, and the solid curves correspond to the total contribution.}
  \label{fig:clas}
\end{figure*}

By exploiting the notation
\begin{align}
\mathcal{C}[w fD] &=x\sum_q e_q^2\int d^2\bm k_T\int d^2 \bm p_T\delta^2(z\bm k_T-\bm P_T+\bm p_T) \nonumber\\
&\times w(\bm k_T, \bm p_T)f^q(x,\bm k_T^2) D^q(z,\bm p_T^2),
\end{align}
we express the structure functions $F_{UU}$ and $F_{LL}^{\cos\phi_h}$ as~\cite{Bacchetta:0611265}
\begin{align}
F_{UU} & = \mathcal{C}[f_1 D_1], \label{eq:FUU}\\
F^{\cos \phi_h}_{LL} & \approx -\frac{2M}{Q} \,\mathcal{C}\,
   \left[\frac{\boldsymbol{\hat{P}_{T}} \cdot \boldsymbol{p_T}}{z M_h}
         \left( \xbj\, e_L H_1^{\perp} \right)\right.\nonumber\\
&\left.+\frac{\boldsymbol{\hat{P}_{T}}\cdot
    \boldsymbol{k_T}}{M}\left(\xbj\, g_L^{\perp} D_1\right)\right].\label{eq:FLL}
\end{align}
Here, we introduce the unit vector $\hat {\bm P}_T={\bP\over P_T}$ and use $M_h$ denote the mass of the final hadron.
As we restrict our scope on the role of the twist-3 distributions in $A_{LL}^{\cos\phi_h}$, in Eq.~(\ref{eq:FLL}) we have suppressed the terms containing the twist-3 fragmentation functions $\tilde{E}$ and $\tilde{D}^\perp$.

The asymmetry $A_{LL}^{\cos\phi_h}$ as a function of $x$ can be cast into
\begin{align}
A_{LL}^{\cos\phi_h}(x)
=&\frac{\int dP_T^2 \int dy \int dz\;\mathcal{C}_{f} \sqrt{2\varepsilon(1-\varepsilon)} \;F_{LL}^{\cos\phi_h}}
{\int d P_T^2 \int dy \int dz \;\mathcal{C}_f \;F_{UU}},\label{eq:asy}
\end{align}
where the kinematical factor $\mathcal{C}_f$ is defined as
\begin{align}
\mathcal{C}_{f}&=\frac{1}{x y Q^2}\frac{y^2}{2(1-\varepsilon)}\Bigl(1+ \frac{\gamma^2}{2x}\Bigl).
\end{align}
In a similar way, we can define the $z$-dependent and the $P_T$-dependent asymmetries.

\begin{figure*}
  \includegraphics[width=0.67\columnwidth]{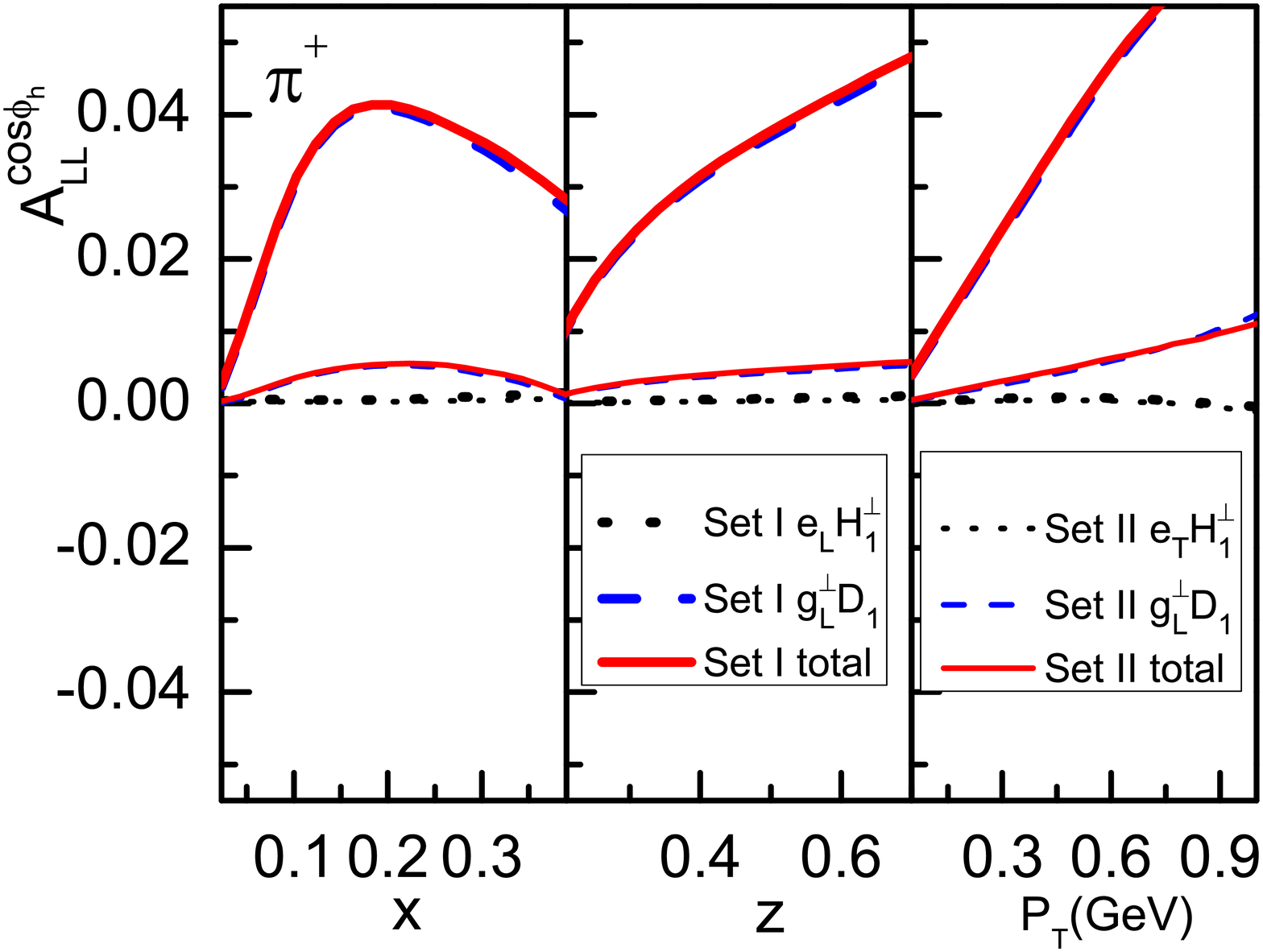}
   \includegraphics[width=0.67\columnwidth]{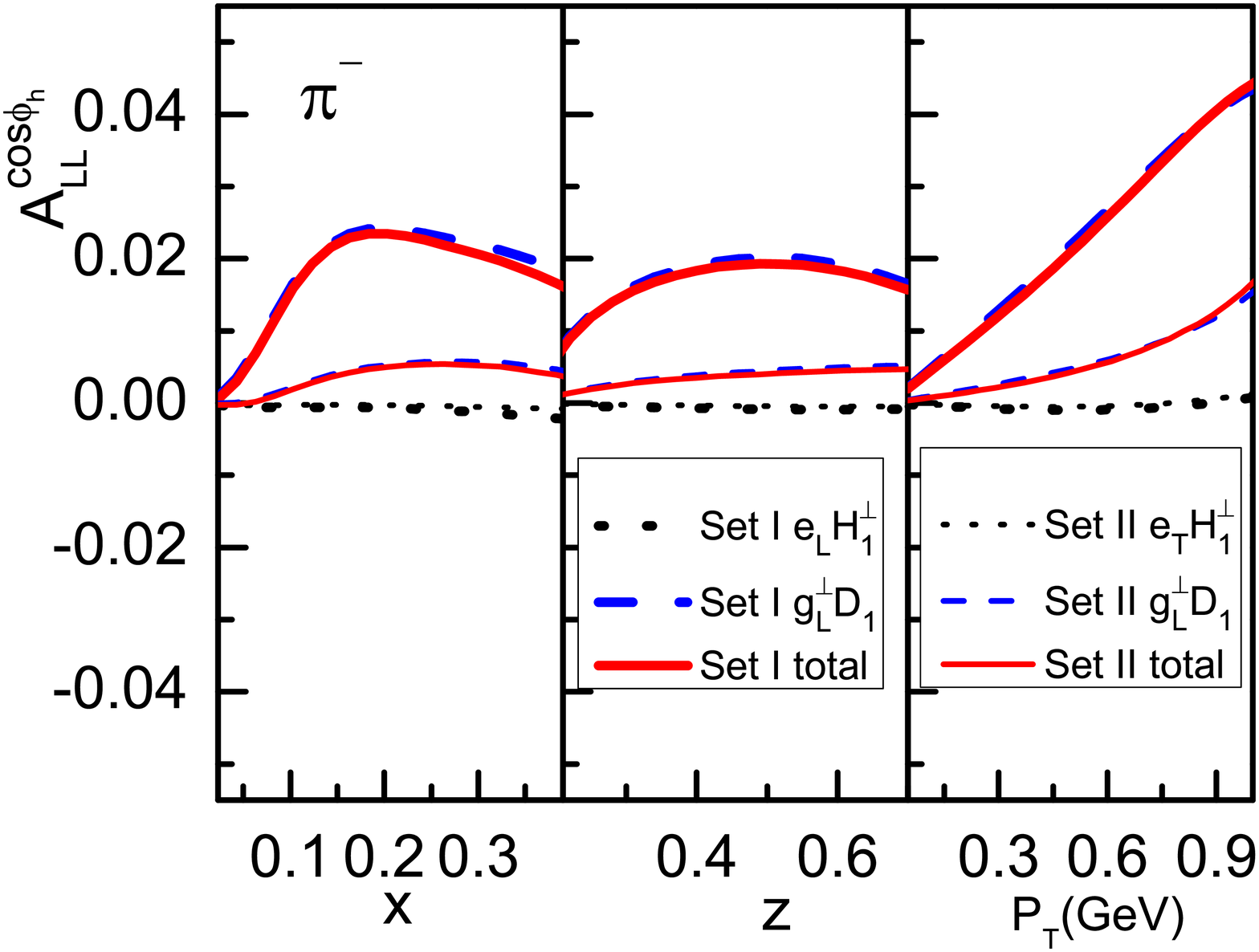}
    \includegraphics[width=0.67\columnwidth]{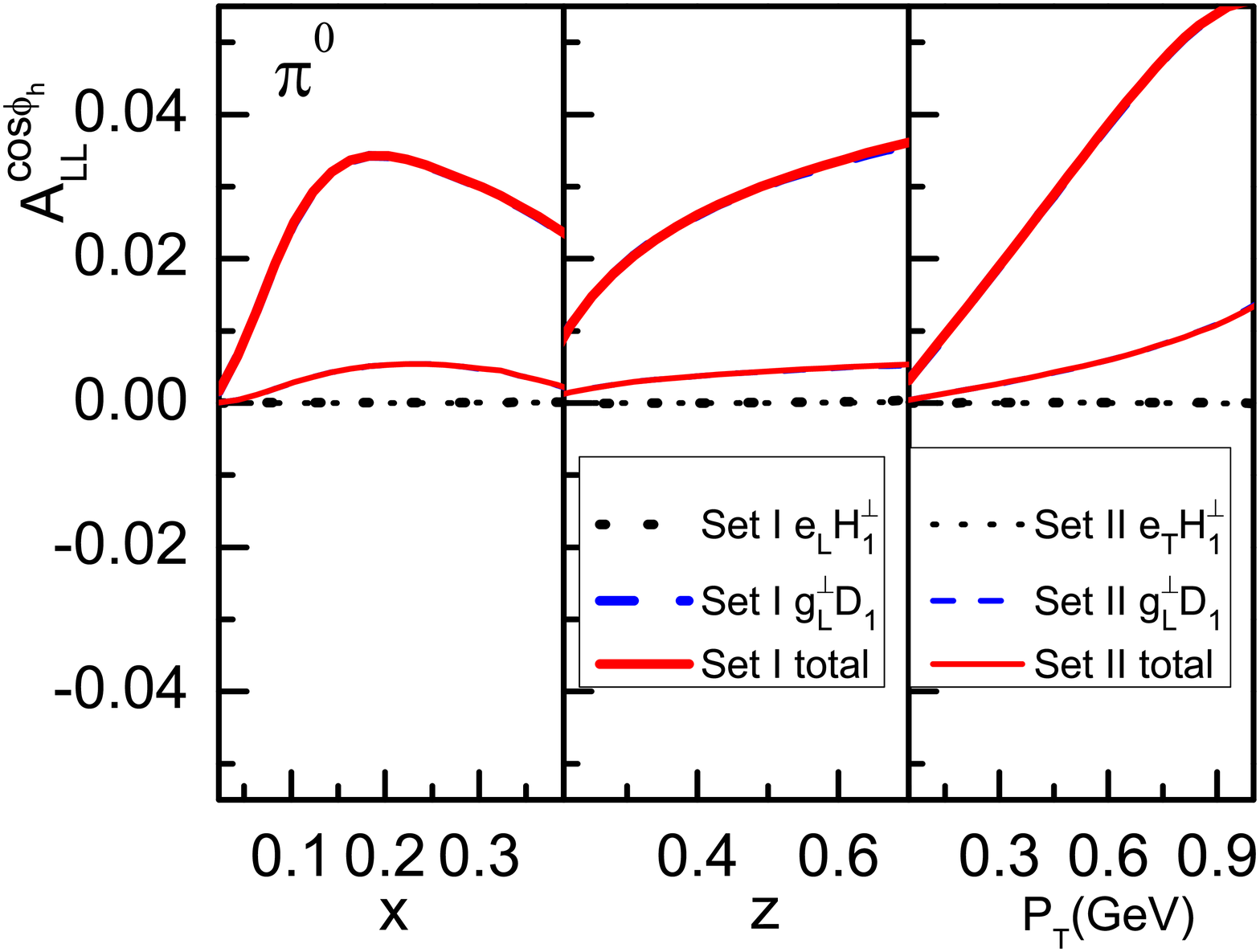}
  \caption{Similar to Fig.~\ref{fig:clas}, but for $A_{\mathrm{LL}}^{\cos\phi_h}$ of $\pi^+$, $\pi^-$, and $\pi^0$ as functions of $x$, $z$ and $P_T$ in SIDIS at HERMES off a proton target.}
  \label{fig:hermes}
\end{figure*}

\begin{figure*}
  \includegraphics[width=0.67\columnwidth]{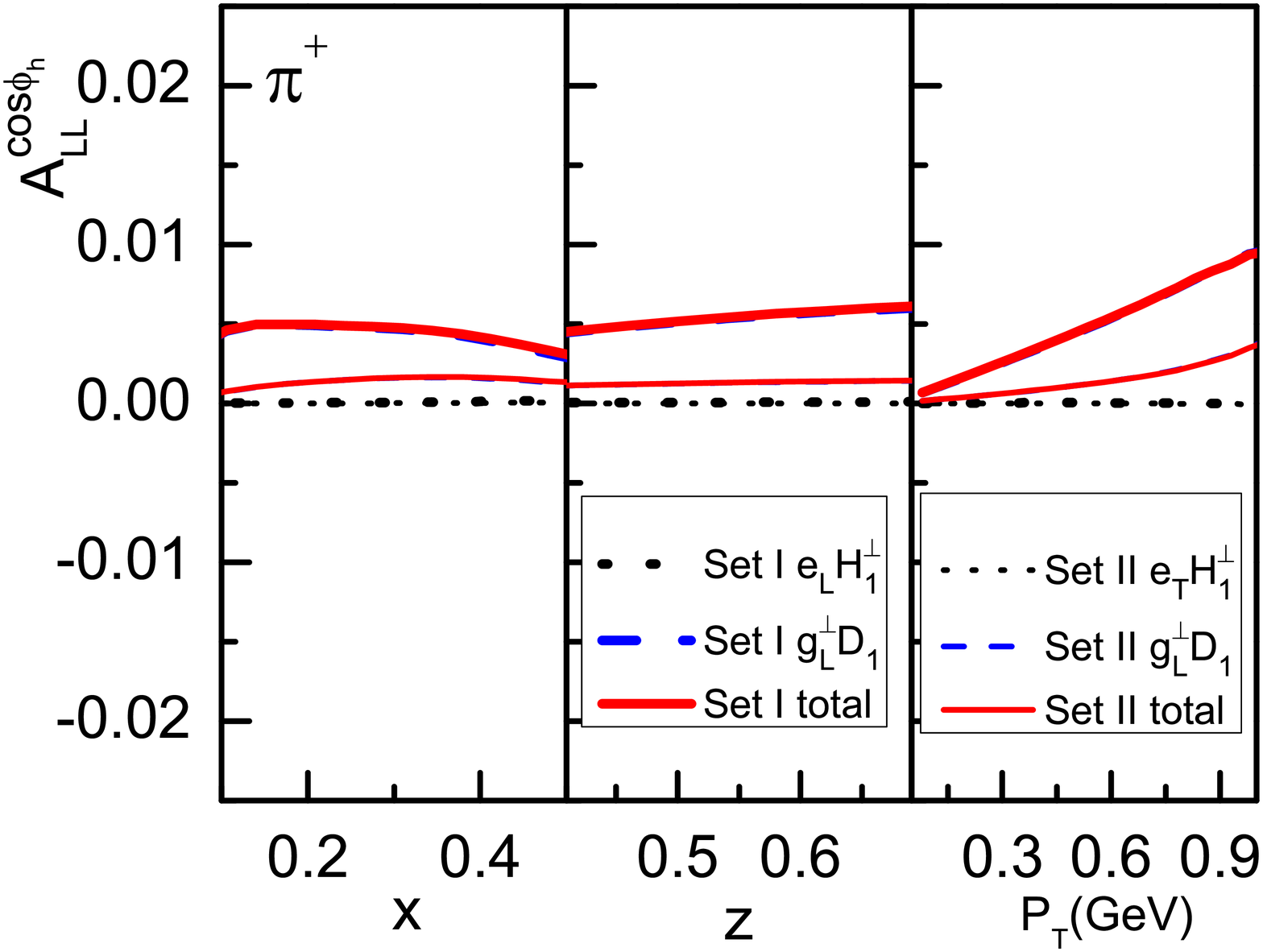}
  \includegraphics[width=0.67\columnwidth]{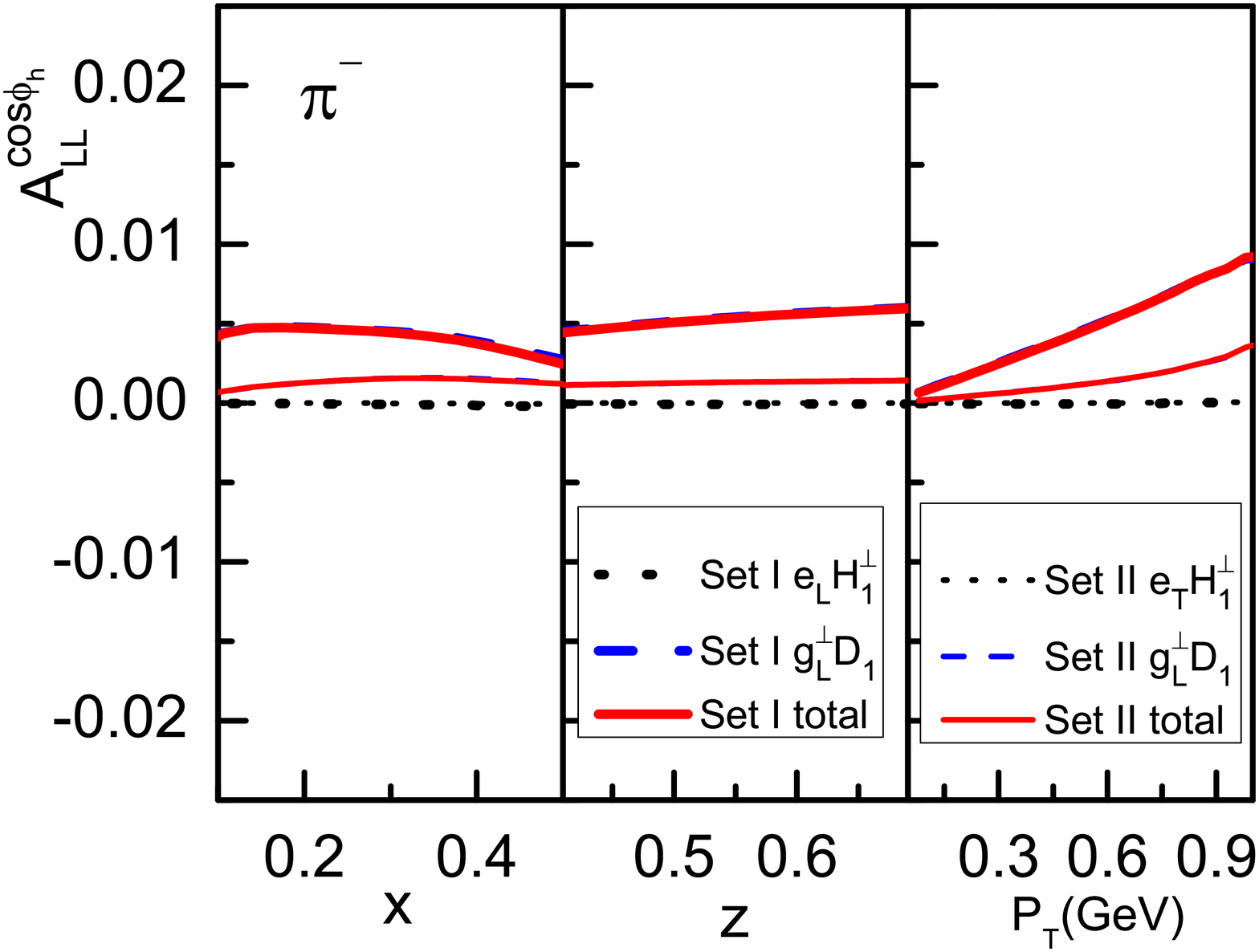}
  \includegraphics[width=0.67\columnwidth]{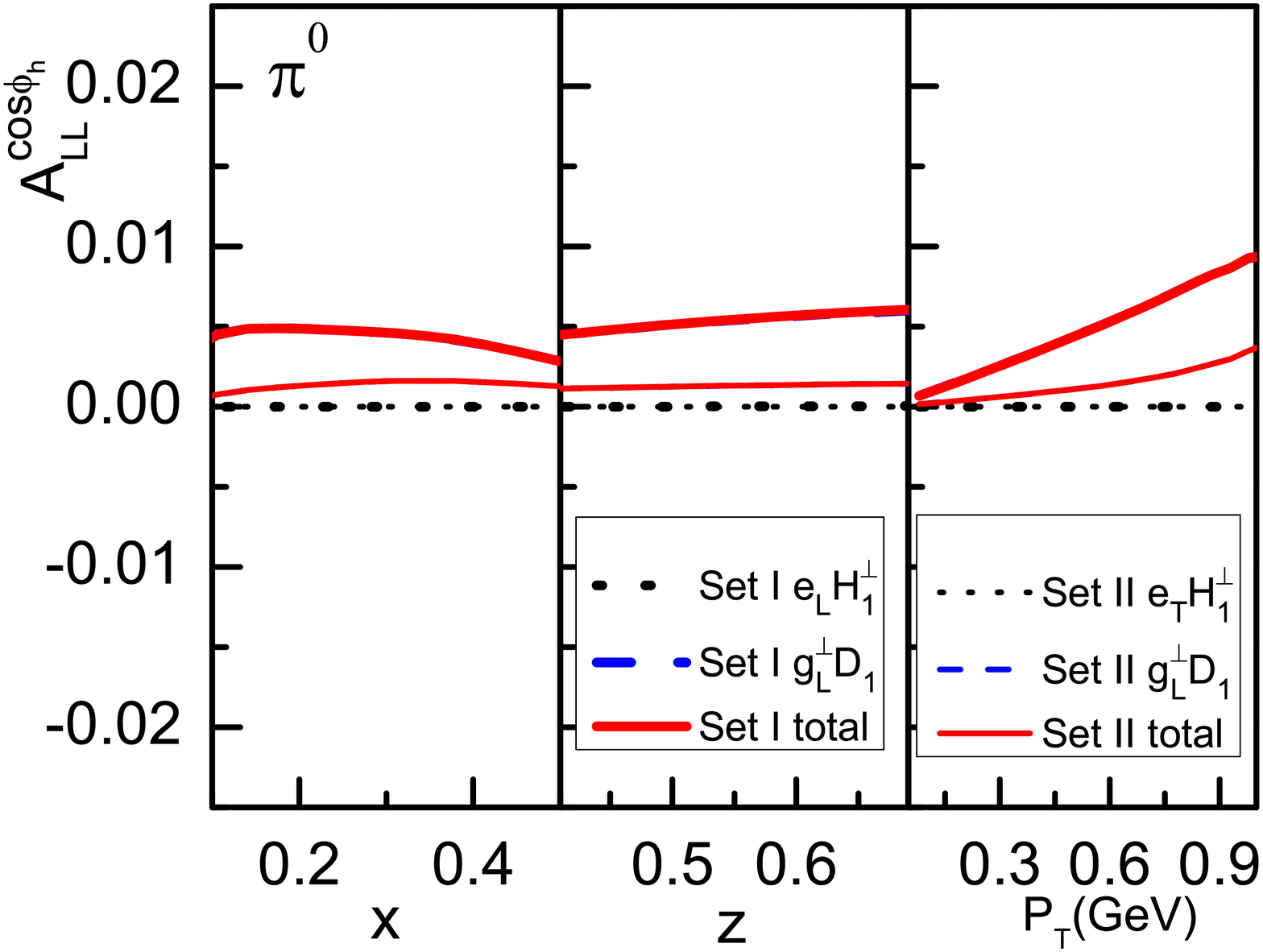}
  \caption{The asymmetry $A_{\mathrm{LL}}^{\cos\phi_h}$ for $\pi^+$, $\pi^-$, and $\pi^0$ as functions of $x$, $z$ and $P_T$ in SIDIS at COMPASS off a deuteron target.}
  \label{fig:compass}
\end{figure*}

In order to estimate the numerical results of $A_{LL}^{\cos\phi_h}$,  we apply the model results of $g_L^\perp$ and $e_L$ obtained in the previous section .
As for the Collins function $H_1^\perp$ appearing in Eq.~(\ref{eq:FLL}), we adopt the parametrization (the standard set) from Ref.~\cite{Anselmino:2013vqa}.
To obtain the Collins function for neutral pion, we employ the following isospin relation:
\begin{align}
 H_1^{\perp \pi^0/u}&=H_1^{\perp \pi^0/d}\equiv{1\over 2}\left( H_{1 \textrm{fav}}^{\perp}+H_{1 \textrm{unf}}^{\perp}\right).
\end{align}
For the TMD fragmentation function $D_1^q\left(z,\bp^2\right)$, which couples with the distribution $g_L^\perp$, we assume its $p_T$ dependence has a Gaussian form
\begin{align}
D_1^q\left(z,\bp^2\right)=D_1^q(z)\, \frac{1}{\pi \langle p_T^2\rangle}
\, e^{-\bm p_T^2/\langle p_T^2\rangle},
\end{align}
where $D_1^q(z)$ is the integrated fragmentation function $D_1^q(z)$, for which we adopt the leading order set of the DSS parametrization~\cite{Florian:2007prd}.
For the Gaussian width $\langle p_T^2\rangle$, we choose its numerical value as $0.2$ \textrm{GeV}$^2$, following the fitted result in Ref.~\cite{Anselmino:2005prd}.
Finally, in our calculation, we consider the kinematical constraints on the intrinsic transverse momentum of the initial quarks given in Ref.~\cite{Boglione:2011}.

We apply the following kinematics to estimate the asymmetry $A_{LL}^{\cos\phi_h}$ at CLAS:
\begin{align}
&E_e =5.5~\textrm{GeV},~~ 0.1<x<0.6,~~ 0.4<z<0.7,\nonumber\\
&Q^2>1\, \textrm{GeV}^2,~~ P_T>0.05\,\textrm{GeV},~~ W^2>4\,\textrm{GeV}^2. \nonumber
\end{align}
In the left, central, and right panels of Fig.~\ref{fig:clas}, we plot the asymmetry for $\pi^+$, $\pi^-$ and $\pi^0$ as functions of $x$, $z$, and $P_T$.
The thick and thin curves correspond to the asymmetries that calculated from the TMD distributions of Set I and Set II, respectively.
The dotted curves show the asymmetries contributed by $e_L$, the dashed curves show those contributed by $g_L^\perp$, while the solid curves denote the total contribution.
We find that the asymmetries calculated from both sets of TMD distributions are positive for all three pions.
Also, It is clear that the asymmetries contributed by the T-even distribution $g_L^\perp$ dominate, and those contributed by $e_L$ are almost negligible.
The asymmetries calculated from Set I TMD distributions are about 5 to 10 percent in magnitude, and are several times larger than those from Set II distributions.

Using the two sets of TMD distributions, we also predict the asymmetry $A_{LL}^{\cos\phi_h}$ at HERMES with a 27.6 GeV positron beam off a proton target~\cite{Airapetian:2005jc}:
\begin{align}
&0.023 < x < 0.4,\,0.1 < y < 0.85,\, 0.2<z<0.7 \nonumber\\
& W^2 > 10\, \textrm{GeV}^2, \, Q^2 >\, 1 \textrm{GeV}^2,\,4\,\textrm{GeV} < P_{h} < 13.8\,\textrm{GeV},\nonumber
\end{align}
and at COMPASS with a 160 GeV muon beam scattered off a deuteron target~\cite{Alekseev:2010dm}:
\begin{align}
&0.004<x<0.7,~~y>0.1,~~0.2<z<0.9,~~ \nonumber\\
&x_F>0,~~Q^2>1\, \textrm{GeV}^2,~~0.1 \textrm{GeV}<P_T<1\,\textrm{GeV},~~ \nonumber \\
 & 5 \,\textrm{GeV}<W<18\,\textrm{GeV}. \nonumber
\end{align}
In Figs.~\ref{fig:hermes} and \ref{fig:compass}, we show our prediction on the $\cos\phi_h$ asymmetry for charged and neutral pions at HERMES and COMPASS, respectively.
We find that the asymmetry at HERMES is smaller than that at CLAS.
Again, the T-even distribution $g_L^\perp$ gives the dominant contribution.
The asymmetries for all pions from the two sets of TMD distributions at COMPASS are consistent with zero (less than 0.5\%).

\section{Conclusion}
\label{conclusion}

In this work, we investigated the $\cos\phi_h$ azimuthal asymmetry in the double longitudinally polarized SIDIS.
Particularly, we focused on the role of the genuine twist-3 TMD distributions and ignored the contribution from the twist-3 fragmentation functions.
To give a quantitative estimate on the $\cos\phi_h$ asymmetry for different pions, we calculated $e_L$ and $g_L^\perp$ of the valence quarks within the framework of spectator model.
We considered two different forms for the propagator of the vector diquark as well as different choices on the flavor separation to obtain two sets of TMD distributions, which were used to predict the asymmetry $A_{LL}^{\cos\phi_h}$ for $\pi^+$, $\pi^-$ and $\pi^0$ at the kinematics of CLAS, HERMES and COMPASS.
We found that the asymmetries from both sets of TMD distributions are sizable at CLAS, while at HERMES only the asymmetry from the distribution of Set I is measurable.
Therefore, it would be feasible to access the $\cos\phi_h$ asymmetry at least at CLAS.
We also found that the predicted asymmetries are positive for all the pions.
This is different from the estimate~\cite{Anselmino:2006yc} based on the Cahn effect, which predicts negative asymmetries for pions, although the magnitude of the asymmetry in our calculation is consistent with the result in Ref.~\cite{Anselmino:2006yc}.
A positive $\cos\phi_h$ asymmetry thus can be viewed as a clear signal of dynamical twist-3 effect that is different from the Cahn effect.
Furthermore, our study shows that the main contribution to the asymmetry is from the T-even distribution $g_L^\perp$, while the contribution from the T-odd distribution $e_L$ almost vanishes.
Future experimental data on the azimuthal asymmetry in the double longitudinally polarized reaction will clarify the role of the twist-3 TMD distributions.

\section*{Acknowledgements}
This work is partially supported by the National Natural Science
Foundation of China (Grants No.~11120101004, No.~11005018, and No.~11035003), and by the Qing Lan Project (China).
W.M. is supported by the Scientific Research Foundation of the Graduate School of SEU (Grant No.~YBJJ1336).

\end{document}